\documentclass{article}
\usepackage[utf8]{inputenc}

\usepackage[authoryear,longnamesfirst]{natbib}

\RequirePackage{amsthm,amsmath,amsfonts,amssymb}
\RequirePackage{graphicx,url}
\usepackage{enumitem,amsmath}   
\usepackage{comment}
\usepackage{algorithm,color}
\usepackage{algpseudocode}

\newtheorem{theorem}{Theorem}[section]

\theoremstyle{remark}
\newtheorem{definition}[theorem]{Definition}
\newtheorem{example}[theorem]{Example}
\newtheorem{conjecture}{Conjecture}

\newtheorem{remark}{Remark}

\newcommand{\RR}{\mathbb{R}}
\newcommand{\R}{\mathbb{R}}

\newcommand{\Tn}{\mathcal{U}_m}

\newcommand{\Trop}{\text{Trop}}

\begin{document}

\title{Tropical Density Estimation of Phylogenetic Trees}
\author{Ruriko Yoshida \and David Barnhill \and  Keiji Miura \and Daniel Howe}
\date{}

\maketitle

\abstract{Much evidence from biological theory and empirical data indicates that, gene tree,  phylogenetic trees reconstructed from different genes (loci), do not have to have exactly the same tree topologies.   Such  incongruence between gene trees might be caused by some ``unusual''  evolutionary events, such as meiotic  sexual  recombination  in eukaryotes or horizontal  transfers  of genetic  material  in prokaryotes. However, most of gene trees are constrained by the tree topology of its species tree, that is, the phylogenetic tree of a given species following their evolutionary history. 
In order to discover ``outlying'' gene trees which do not follow the ``main distribution(s)'' of trees, we propose to apply the ``tropical metric'' with the max-plus algebra from tropical geometry to a non-parametric estimation of gene trees over the space of phylogenetic trees.
In this research we apply the ``tropical metric,'' a well-defined metric over the space of phylogenetic trees under the max-plus algebra, to non-parametric estimation of gene trees distribution over the tree space. Kernel density estimator (KDE) is one of the most popular non-parametric estimation of a distribution from a given sample, and we propose an analogue of the classical KDE in the setting of tropical geometry with the tropical metric which measures the length of an intrinsic geodesic between trees over the tree space.  We estimate the probability of an observed tree by empirical frequencies of nearby trees, with the level of influence determined by the tropical metric.
Then, with simulated data generated from the multispecies coalescent model, we show that the non-parametric estimation of gene tree distribution using the tropical metric performs better than one using the Billera-Holmes-Vogtmann (BHV) metric developed by Weyenberg et al.~in terms of computational times and accuracy. We then apply it to Apicomplexa data.
}

\section{Introduction}

Due to efficient genome sequencing technologies in terms of time and cost, it is essential to develop efficient bioinformatic methods to analyze genome structure and evolution. In this paper, we focus on correlations between {\em gene trees}, that is, phylogenetic trees reconstructed from alignments of genes in a genome. An\'e et al., for example, applied a Bayesian method to estimate concordance among gene trees from multiple loci in \cite{Ane2007}.  However, most of such estimators assume that concordance present among the given set of gene trees.  In practice, however, numerous evolutionary processes can reduce correlations between gene trees. For example, such evolutionary processes include negative or balancing selection on a locus, which might increase the chance for ancestral gene copies to maintain through speciation events \cite{Takahata}, and horizontal gene transfer, which shuffles divergent genes among different species \cite{Liu2007}.  

In this paper, we propose a method to estimate a distribution of gene trees over the {\em space of phylogenetic trees} as a whole. Especially using this estimated distribution of trees, our interest is to identify gene trees which exhibit significant discordance among gene trees.  These ``uncommon''  genes can be generated from evolutionary processes, such as, paralogy, neofunctionalization, horizontal gene transfer or periods of rapid molecular evolution, and they might come from processes of data analyses, such as incorrect sequencing, alignment, tree reconstruction or annotation \cite{Horner}.

While parametric statistical methods are available or under development, we propose in this paper a {\em nonparametric} approach which offers particular advantages in phylogenomic analyses. In particular, problems of estimation and potentially incorrect selection of model parameters, which could obscure the search for outlier trees, are obviated. Here we propose a method analogue to kernel density estimator to estimate a distribution of trees over the space of trees using tools from tropical geometry.

One of the most challenging problems in phylogenomics is to study correlations among gene trees over the space of phylogenetic trees.  
Ideally, we could apply conventional  statistical methods directly to a sample of gene trees, however, the {\em space of phylogenetic trees}, the space of all possible phylogenetic trees with $m$ leaves, is not Euclidean.
In fact it is an union of lower dimensional cones over $\mathbb{R}^e$, where $e = \binom{m}{2}$ and it is not convex \cite{AK}.
Therefore, we cannot just apply  conventional statistical models in data science to a set of phylogenetic trees because these methods assume Euclidean spaces \cite{YZZ}.

The notion of the space of phylogenetic trees with $m$ leaves comes from the work by  Billera-Holmes-Vogtmann (BHV) in \cite{BHV}.  Billera-Holmes-Vogtmann defined the space by gluing $m-2$ dimensional positive orthants, where each orthant represents all possible rooted phylogenetic trees with a fixed tree topology with $m$ leaves.
Over the tree space with the {\em BHV metric}, two orthants with coordinates defined by edge lengths of interior edges are glued to each other if the tree topology for one orthant differs by one nearest neighbor interchange (NNI) distance to the tree topology for the other orthant.
In the work, Billera-Holmes-Vogtmann also showed that this tree space is ${\rm CAT}(0)$ space.
This means that for any two trees in the tree space there is a unique shortest connecting path, called a geodesic, defined by the ${\rm CAT}(0)$-metric.

Shortly after that, in 2004, Speyer and Sturmfels showed that the space of phylogenetic trees with $m$ leaves is a tropical Grassmanian \cite{SS}, which is a {\em tropicalization} of the set of all solutions for a system of certain linear equations \cite{YZZ} under max-plus arithmetic.
Several researchers then showed that the {\em tropical metric} with max-plus algebra on the space of {\em equidistant trees} with $m$ leaves 
behaves very well \cite{AGNS,CGQ,LSTY}.
For example, the dimension of the convex hull, the smallest tropical convex set, of $s$ points with the tropical metric over the {\em tropical projective space} $(\mathbb{R} \cup \{-\infty\})^e/\mathbb{R}{\bf 1}$ is at most $s-1$ while this is not the case with the BHV metric \cite{LSTY}.
Therefore developing a machine learning algorithm that predicts based on the tropical metric as a data proximity measure is crucial for novel discovery.

Applications of the tropical metric to phylogenomics have been often done over the space of equidistant trees. 
An equidistant tree is a rooted phylogenetic tree whose distance between its root to each leaf is the same for all leaves in the tree.
In terms of biology, this can be seen as a phylogenetic tree with a molecular clock. Also, the multispecies coalescent model assumes that all gene trees are equidistant.
Therefore this is a natural assumption in evolutionary biology \cite{mesquite}.
For example, Yoshida et al.~\cite{YZZ} and Page et al.~\cite{10.1093/bioinformatics/btaa564} developed {\em tropical principal component analysis}.
In their work they use the fact that the space of equidistant trees with $m$ leaves is a tropically convex set over the tropical projective space in terms of the tropical metric and the {\em tropical line segment} between any trees over the space is intrinsically geodesic and is unique \cite{MLKY}.  

A kernel density estimator (KDE) is a non-parametric density estimator using kernel functions,
which is useful for, say, discovering outliers.
Weyenberg et al.~developed a non-parametric density estimator 
over the space of phylogenetic trees in terms of the BHV metric by mimicking a classical KDE \cite{KDE}.
The biggest problem Weyenberg et al.~encountered was that the kernel function normalizing constant varies depending on the location of the center of the function. In addition, even though Weyenberg et al.~developed a method to approximate the normalizing constant for a kernel function with the BHV metric over the space of phylogenetic trees, there is still no explicit method to compute the normalizing constant.

In this paper, since the space of equidistant trees is a tropical convex set \cite{YZZ}, we apply a Hit and Run (HAR) sampler from tropically convex sets with the tropical metric developed by Yoshida et al.~\cite{YMB} to estimate the normalizing constant of a kernel function with the tropical metric over the space of equidistant trees. Computationally, we show that the normalizing constant of a kernel function is independent from a central location of the function over the space of equidistant trees.
Then we develop an analogue of a classical KDE with the tropical metric over the space of equidistant trees and, with simulated data generated from the multispecies coalescent model, we show that the KDE with the tropical metric performs better than one with the BHV developed by Weyenberg et al.~\cite{KDETree} in terms of computational time and accuracy.
We also apply it to Apicomplexa data from \cite{kuo}.

This paper is organized as follows.
In Section \ref{sec:basics}, we first outline basics on tropical geometry using the max-plus algebra over the tropical semiring. Then, we outline our non-parametric estimation of gene tree distribution over the space of phylogenetic trees with a given set of leaves defined by the tropical metric.  In Section \ref{sec:result}, we show how we set up simulation studies with our method.  Then we show the results from computational experiments with simulated data generated from the multispecies coalescent model and with the empirical data of Apicomplexa from \cite{kuo}.  In Section \ref{sec:discussion}, we discuss the results from computational experiments and we end with future work and an open problem in Section \ref{sec:conclusion}.

\section{Methods}\label{sec:basics}

\subsection{Basics of Tropical Geometry}

Throughout this paper, like \cite{SS}, we consider the {\it tropical projective torus} $\mathbb R^e \!/\mathbb R {\bf 1}$, which is isomorphic to $\R^{e-1}$.
For more details, see \cite{ETC,MS}.

\begin{definition}[Tropical Arithmetic Operations]
Under the tropical semiring $(\,\mathbb{R} \cup \{-\infty\},\boxplus,\odot)\,$, the tropical arithmetic operations of addition and multiplication are defined as:
$$c_1 \boxplus c_2 := \max\{c_1, c_2\}, ~~~~ c_1 \odot c_2 := c_1 + c_2,$$
where $ c_1, \, c_2 \in \mathbb{R}\cup\{-\infty\}.$
Over the tropical semiring, the identity element under addition is $-\infty$ and the identity element under multiplication is $0$.
\end{definition}

\begin{definition}[Tropical Scalar Multiplication and Vector Addition]
For any scalars $c_1,\, c_2 \in \mathbb{R}\cup \{-\infty\}$ and for any vectors $v = (v_1, \ldots ,v_e), w= (w_1, \ldots , w_e)$ over the {\em tropical projective space} $(\mathbb{R}\cup-\{\infty\})^e\!/\mathbb R {\bf 1}$, we have tropical scalar multiplication and tropical vector addition as:
$$c_1 \odot v \boxplus c_2 \odot w := (\max\{c_1+v_1,c_2+w_1\}, \ldots, \max\{c_1+v_e,c_2+w_e\}).$$
\end{definition}

\begin{definition}\label{def:polytope}
Suppose we have $S \subset \mathbb R^e \!/\mathbb R {\bf 1}$. 
$S$ is {\em tropically convex} if
\[
c_1 \odot v \boxplus c_2 \odot w \in S
\]
for any $c_1, c_2 \in \R$ and for any points $v, w \in S$. 
Suppose $V = \{v^1, \ldots , v^s\}\subset \mathbb R^e \!/\mathbb R {\bf   1}$.  The smallest tropically-convex subset containing $V$ is called the {\em tropical convex hull} or {\em tropical polytope} of $V$ which can be written as the set of all tropical linear combinations of $V$ as:
$$ \mathrm{tconv}(V) = \{a_1 \odot v^1 \oplus a_2 \odot v^2 \oplus \cdots \oplus a_s \odot v^s \mid  a_1,\ldots,a_s \in \R \}.$$
A {\em tropical line segment} between two points $v^1, \, v^2$ is a tropical polytope of a set of two points $\{v^1, \, v^2\} \subset \mathbb R^e \!/\mathbb R {\bf   1}$.
\end{definition}

\begin{definition}[Generalized Hilbert Projective Metric]
\label{eq:tropmetric} 
For any vectors $v:=(v_1, \ldots , v_e), \, w := (w_1, \ldots , w_e) \in \mathbb R^e \!/\mathbb R {\bf 1}$,  the {\em tropical distance} $d_{\rm tr}$ between $v$ and $w$ is defined as:
\begin{equation*}
d_{\rm tr}(v,w)  := \max_{i \in \{1, \ldots , e\}} \bigl\{ v_i - w_i \bigr\} - \min_{i \in \{1, \ldots , e\}} \bigl\{ v_i - w_i \bigr\}.
\end{equation*}
This distance measure is a well-defined metric over the tropical projective torus $\mathbb R^e \!/\mathbb R {\bf 1}$ \cite{LSTY}. 
\end{definition}

\subsection{Basics of Ultrametrics}

Suppose we have $[m] := \{1, \ldots , m\}$ and let $d: [m] \times [m] \to \RR$ be a metric over $[m]$, that is, $d$ is a map from $[m]\times [m]$ to $\RR$ such that
\begin{eqnarray}\nonumber
d(i, j) = d(j, i) & \mbox{for all } i, j \in [m]\\\nonumber
d(i, j) = 0 & \mbox{if and only if } i = j\\\nonumber
d(i, j) \leq d(i, k) + d(j, k) & \mbox{for all }i, j, k \in [m].
\end{eqnarray}

Suppose $d$ is a metric on $[m]$.  Then if 
\begin{eqnarray}
\max\{d(i, j), d(i, k), d(j, k)\} 
\end{eqnarray}
is attained at least twice for any $i,j,k \in [m]$, then $d$ is called an {\em ultrametric}.  

\begin{example}
Suppose $m = 3$.  Let $d$ be a metric on $[m]:=\{1, 2, 3\}$ such that
\[
d(1, 2) = 2, \, d(1, 3) = 2,\, d(2, 3) = 1.
\]
Since the maximum is achieved twice, $d$ is an ultrametric.
\end{example}

A phylogenetic tree is a weighted tree whose internal nodes do not have labels and whose external nodes, i.e., leaves, have labels $[m]$.  Throughout this paper, we consider a rooted phylogenetic tree with a leaf label set $[m]$.  
\begin{definition}
Suppose we have a rooted phylogenetic tree $T$ with a leaf label set $[m]$.  If the distance from its root to each leaf $i \in [m]$ is the same distance for all $i \in [m]$, then we call $T$ an {\em equidistant tree}.
\end{definition}

In order to conduct a statistical analysis, we need to map a phylogenetic tree on $[m]$ to a vector representation.  There are many ways to map a phylogenetic tree to a vector, including the BHV coordinates \cite{BHV}.  In this paper, we vectorize phylogenetic tree as dissimilarity maps. Dissimilarity maps are maps $d: [m] \times [m] \to \RR$ such that $d(i, i) = 0$ and $d(i, j) = d(j, i)$.  In phylogenetics, we consider dissimilarity maps over the product of a leaf set $[m]$ such that $d(i, j)$ is the pairwise distance between a leaf $i \in [m]$ to a leaf $j \in [m]$.  Throughout this paper we consider a vector of all possible pairwise distances in $T$ between any two leaves in $[m]$ as a vector representation of a phylogenetic tree $T$ with $[m]$. Then we have the following theorem.

\begin{theorem}[\cite{Buneman}]\label{thm:3pt}
Suppose we have an equidistant tree $T$ with a leaf label set $[m]$ and suppose $d(i, j)$ for all $i, j \in [m]$ is the distance from a leaf $i$ to a leaf $j$.  Then, $d$ is an ultrametric if and only if $T$ is an equidistant tree. 
\end{theorem}

\begin{example}
Suppose we have $m = 5$.
Then, the phylogenetic tree shown in Fig.~\ref{fig:eqEx} is an equidistant tree with a leaf label set $[5]:=\{A, B, C, D, E\}$ and its pairwise distances are 
\[
u = (4, 4, 4, 4, 2, 2, 2, 1.6, 1.6, 0.6)
\]
which is an ultrametric.
\begin{figure}[t]
    \centering
    \includegraphics[width=0.25\textwidth]{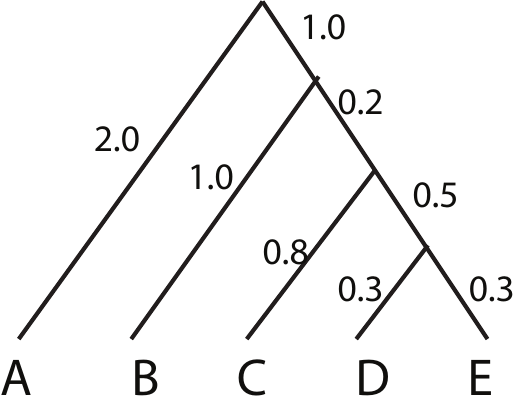}
    \caption{Example of an equidistant tree with a leaf label set $[5]$.}
    \label{fig:eqEx}
\end{figure}
\end{example}

Using Theorem \ref{thm:3pt}, if we wish to consider all possible equidistant trees, then it is equivalent to consider the space of ultrametrics as the space of phylogenetic trees on $[m]$.  Here we define $\Tn$ as the space of ultrametrics with a set of leaf labels $[m]$.

Throughout this paper, we assume we have a sample of gene trees which are equidistant. This assumption is not unusual in phylogenomics since the multispecies coalescent model assumes that all gene trees are equidistant trees in order to conduct the inference on the species tree from a sample of gene trees \cite{mesquite}.

\begin{theorem}[\cite{AK,10.1093/bioinformatics/btaa564}]
Suppose we have a {\em classical} linear subspace $L_m \subset \RR^e$ defined by the linear
equations $x_{ij} - x_{ik} + x_{jk}=0$ for $1\leq i < j <k \leq m$. Let $\mbox{Trop}(L_m)\subseteq \RR^e/\RR {\bf 1}$ be the {\em tropicalization} of the linear space $L_m \subset \RR^e$, that is, we replace the classical addition by the tropical addition $\boxplus$ and we replace the classical multiplication by the tropical multiplication $\odot$ in the equations defining the linear subspace $L_m$, so that all points $(x_{12},x_{13},\ldots, x_{m-1,m})$ in $\Trop(L_m)$ satisfy the condition:
\[
\max_{i,j,k\in [m]}\{v_{ij},v_{ik},v_{jk}\}
\]
is achieved at least twice. 
Then the image of $\mathcal U_m$ inside of the tropical projective torus $\RR^e/\RR {\bf 1}$ is equal to $\Trop(L_m)$. 
\end{theorem}

\begin{remark}
Since $\Tn \subseteq \RR^e/\RR {\bf 1}$ is the tropicalization of the linear subspace, $\Tn$ is tropically convex.  Therefore, if we take a tropical line segment $\Gamma_{u, v}$ between any two ultrametrics $u, v \in \Tn$, then since $\Gamma_{u, v}$ is also tropically convex, $\Gamma_{u, v}$ is contained in $\Tn$, i.e., $\Gamma_{u, v} \subset \Tn$.  Further, Monod et al.~in \cite{MLKY} showed that $\Gamma_{u, v}$ is a unique geodesic between $u, v \in \Tn$.  Therefore, $d_{\rm tr}(u, v)$ measures the length of $\Gamma_{u, v}$ which is an {\em intrinsic metric} between $u, v \in \Tn$.
\end{remark}

\subsection{Non-parametric Estimation of Gene Tree Distribution}

Suppose we have an i.i.d. sample of trees $\mathcal{S}:= \{T_1, \ldots, T_N \}\subset \Tn$. Our goal is to estimate the gene tree distribution from $\mathcal{S}$ over the space of ultrametrics $\Tn$. Here we assume that the 'non-outlying trees’ are independently sampled from some unknown distribution which we are interested in estimating and ’outlying trees’ sampled from a different distribution. 
Our non-parametric density estimator with the tropical metric over the space of ultrametrics $\Tn$ mimics a classical kernel density estimator (KDE)  formulated as:
\begin{equation}\label{mastereq}
\hat{f}(T) \propto \frac{1}{N} \sum_{i=1}^N k(T, T_i)
\end{equation}
where $k$ is a non-negative function defined over $\Tn$ such that
\begin{equation}\label{eq:kernel}
    k(T, T_i) = \exp \left( { - {\left({\frac{d_{\rm tr}(T,T_i)}{\sigma}}\right)}}\right) ,
\end{equation}
where  $\sigma > 0$ is a user specified parameter to define ``bandwidth'' which controls, how tightly each contribution of a function $k(T,T_i)$ will be centered around $T_i \in \mathcal{S}$ in terms of $d_{\rm tr}$ (See Formula \eqref{mastereq}).    In \cite{KDE, KDETree}, the default set up of this user-defined parameter is determined by the nearest neighbor of each $T_i \in \mathcal{S}$.  
Ideally the normalizing constant
\[
C(T_i) = \int_{\Tn} k(T, T_i) dT
\]
does not depend on $T_i \in \Tn$ so that we do not have to compute the normalizing constant for each $T_i \in \mathcal{S}$ as was required in \cite{KDETree}.  In achieving this, our proposed method in this section will be more analogous to a kernel density estimation.  
Since our experiments in the following subsection show that the normalizing constant $C(T_i)$ does not vary for any $T_i \in \Tn$, we assume that the normalizing constant $C(T_i)$ is a constant for any $T_i \in \Tn$.

In this paper we are interested in detecting outliers $T_j \in \mathcal{S}$ similar to \cite{KDE,KDETree}.  Therefore, we consider the estimation
\[
\hat{g}(T_j) \propto \frac{1}{N-1} \sum_{i \neq j} k(T_j, T_i)
\]
for $T_j \in \mathcal{S}$.
As is the case in \cite{KDE,KDETree}, after we estimated $\hat{g}(T_j)$ for each $T_j \in \mathcal{S}$, we classify $T_j$ as an outlying tree if $\hat{g}(T_j)$  is less than $Q_1 - \kappa IQR$, where $Q_1$ is the first quartile and $IQR$ is the interquartile range of the set of all scores for all trees in $\mathcal{S}$. $\kappa$ is a tuning parameter and it is set to 1.5 as a default \cite{Tukey}.

\subsection{Approximating Normalizing Constants}

In \cite{KDETree, KDE}, the authors considered the function
\begin{equation}\label{eq:kernelBHV}
    k_{\rm BHV}(T, T_i) \propto \exp \left( { - {\left({\frac{d_{\rm BHV}(T,T_i)^2}{\sigma}}\right)}}\right) ,
\end{equation}
where $d_{\rm BHV}$ is a BHV metric defined by Billera, Holmes and Vogtmann over $\mathcal{T}_m$, the space of phylogenetic trees with $m$ leaves using the BHV metric \cite{BHV}.  In \cite{KDETree}, Weyenberg et al.~showed that 
$C_{\rm BHV}(T_i)$ varies on $T_i \in \mathcal{T}_m$ where
\[
C_{\rm BHV}(T_i) = \int_{\mathcal{T}_m} k_{\rm BHV}(T, T_i) dT.
\]
Therefore, Weyenberg et al.~in \cite{KDE} developed an algorithm to approximate $C_{\rm BHV}(T_i)$ for any $T_i \in \mathcal{T}_m$. When $T_i$ is the star tree, i.e., the tree with no internal branch, $C_{\rm BHV}(T_i)$ achieves its largest values.  
Therefore, in this section, we apply a {\em Hit and Run} sampler developed by Yoshida et al.~\cite{YMB} to approximate the normalizing constant of $k(T, T_i)$ for $T_i \in \Tn$.  Especially, we compare the normalizing constant  of $k(T, T_i)$ where $T_i$ is the star tree and $T_i$ is a binary random tree for $m = 10$.

\begin{figure*}[t]
    \centering
    \includegraphics[width=0.42\textwidth]{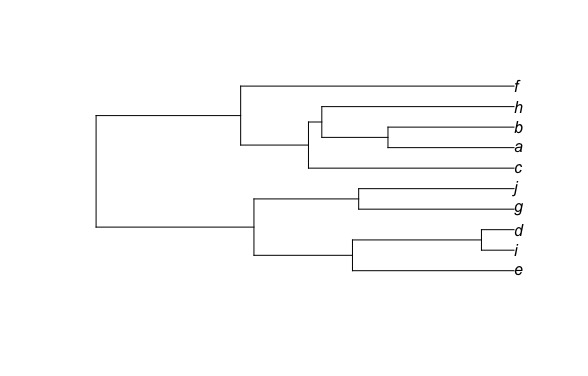}
    \includegraphics[width=0.42\textwidth]{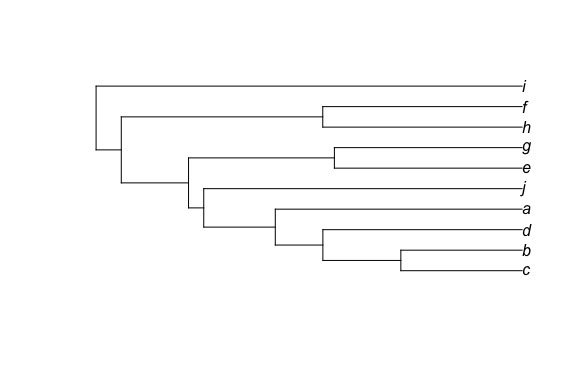}
    \caption{(Left) The centroid tree used for computational experiments. Its ultrametric is
$ u=(0,$ $ 0.446,$ $ 2,$ $ 2,$ $ 0.942,$ $ 2,$ $ 0.348,$ $ 2,$ $ 2,$ $ 0.446,$ $ 2,$ $ 2,$ $ 0.942,$ $ 2,$ $ 0.348,$ $ 2,$ $ 2,$ $ 2,$ $ 2,$ $ 0.942,$ $2,$ $ 0.446,$ $ 2,$ $ 2,$ $ 0.124,$ $ 2,$ $ 0.845,$ $ 2,$ $ 0,$ $ 0.845,$ $ 2,$ $ 0.845,$ $ 2,$ $ 0.124,$ $ 0.845,$ $ 2,$ $ 0.942,$ $ 2,$ $2,$ $ 2,$ $ 0.845,$ $ 0.079,$ $ 2,$ $ 2,$ $ 0.845)$. (Right) The second centroid tree for Example \ref{eg:normalizedconst}. The picture is produced by the {\tt R} package {\tt ape} \cite{APE}.}
    \label{fig:mu}
\end{figure*}

\begin{example}\label{eg:normalizedconst}
In this example, we use three different trees as the centroid of each distribution.  One is shown in the left picture of Fig.~\ref{fig:mu}.  The second is shown in the right picture of Fig.~\ref{fig:mu} and the last is the star tree of ten leaves with the length of each branch equal to $1$.   Using 1,000 samples and letting $\sigma = 1.5, 2, 5$ we achieve the results shown in Table \ref{tb:normalizedconst}.   From this result, it seems that the normalized constant $D(T_i)$ in terms of $d_{\rm tr}$ for any $T_i \in \Tn$ is invariant under the different tree topologies. 

\begin{table}{Estimating the normalizing constants}
    \centering
    \begin{tabular}{|c|ccc|}
    \hline
         $\sigma \backslash$ Tree Type&Tree in left Fig.~\ref{fig:mu} &Tree in right Fig.~\ref{fig:mu} & Star Tree \\\hline
         1.5&117.21 & 120.59 & 117.78 \\
         2& 199.24 & 199.24 & 201.23\\
         5& 521.23 & 521.12 & 524.45\\\hline
    \end{tabular}
    \caption{Results for estimating the normalizing constants for different centroids with varying $\sigma$.}
    \label{tb:normalizedconst}
\end{table}

\end{example}

\subsection{Computational time}

The computational time complexity of the tropical distance $d_{\rm tr}(T_1, T_2)$ between two trees $T_1, T_2 \in \Tn$ is $O(m^2)$.  Therefore, when computing the normalizing constant, for each $T \in \Tn$, the time complexity of computing $f(T)$ is $O(Nm^2)$ while with the BHV metric it is $O(Nm^6)$ for each $T \in \mathcal{T}_m$. 
\section{Results}\label{sec:result}

\subsection{Simulated Experiments}

For this computational experiment, we generate gene trees from the multispecies coalescent models with a given species tree via the software {\tt Mesquite} \cite{mesquite}.  We fixed
the effective population size $N_e = 100,000$ and varied
$R = \frac{SD}{N_e}$
where $SD$ is the species depth which is the number of generations from the common ancestor (the root) to the taxa (leaves). 

\begin{algorithm}
\caption{Generating a set of gene trees from the multispecies coalescent model}\label{alg:sim0}
\begin{algorithmic}
\State {\bf Input:} The number of leaves $m$; $R$, the ratio of the species depth and effective population size; and the number of gene trees $N$.
\State {\bf Output:} A sample of gene trees $\mathbb{T}$.
\State Set the labels for leaves to the species tree and gene trees using $m$.
\State Use the Yule model to generate a random species tree $T$.
\State Using the species tree $T$ with the ratio $R$, generate $N$ gene trees $\mathbb{T}$.
\Return $\mathbb{T}$.
\end{algorithmic}
\end{algorithm}

To sample trees randomly from two different distributions, 
we fix
the number of leaves as
$m = 10$ and generate two different species trees $T_1, \, T_2$ using the Yule process.  Then using the coalescent model for gene trees within the species tree, we generate $1000$ gene trees for each species tree via Algorithm \ref{alg:sim0}.  In these simulated
experiments, we vary the ratio $R = 0.25, 0.5, \, 
1, \, 2, \, 5, \, 10$.
Let $\mathbb{T}_1$ be the set of gene trees with the species tree $T_1$ and let $\mathbb{T}_2$ be the set of gene trees with the species tree $T_2$.  It is worth noting that when we have small $R$, gene trees generated from a coalescent model within a given species tree are similar to random trees.  Thus, it becomes harder to distinguish between two distributions of gene trees with two different species trees as $R$ becomes smaller \cite{rannala:hal-02535622}.  

\begin{algorithm}
\caption{Experiments on a Sample Generating from Coalescent}\label{alg:sim1}
\begin{algorithmic}
\State {\bf Input:} $g > 1$ many non-outlier gene trees $T_1, \ldots , T_g$; and $r \geq 1$ many outlier gene trees $T'_1, \ldots , T'_r$.  Density Estimator $M$.
\State {\bf Output:} Estimated probabilities for $g$ many non-outlier gene trees and $r$ many outlier gene trees.
\For{$j= 1, \ldots , r$,}
\For{$i= 1, \ldots , g$,}
\State Compute estimated probability $\hat{f}(T_i)$ of $T_i$ via $M$ with a sample of gene trees $\{T_1, \ldots , T_{i-1}, T_{i+1}, \ldots , T_g, T'_j\}$. 
\State Compute estimated probability $\hat{f}(T'_j)$ of $T'_j$ via $M$ with a sample of gene trees $\{T_1, \ldots , T_g\}$. 
\EndFor 
\EndFor \\
\Return $\hat{f}(T_1), \ldots , \hat{f}(T_g)$ and $\hat{f}(T'_1), \ldots , \hat{f}(T'_r)$.
\end{algorithmic}
\end{algorithm}

\begin{figure*}
    \centering
    \includegraphics[width=0.475\textwidth]{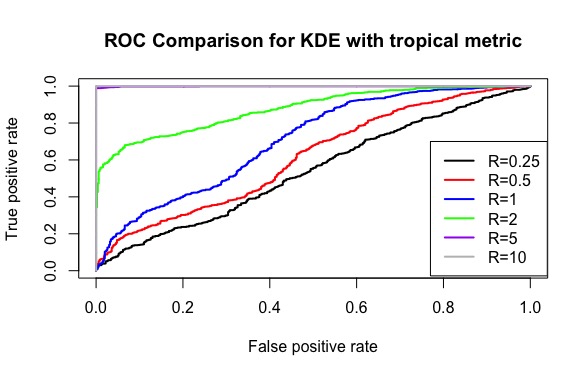}
    \includegraphics[width=0.475\textwidth]{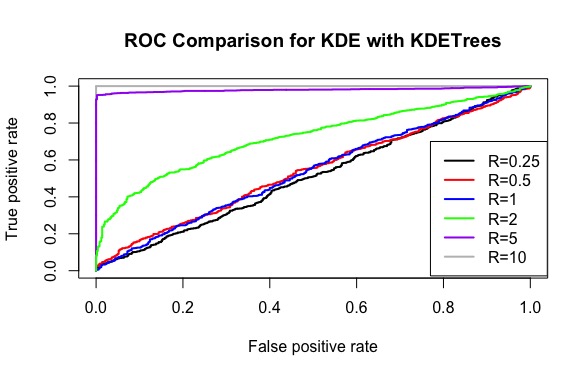}
    \caption{(Left) ROC curves for the KDE with the tropical metric.  (Right) ROC curves for {\tt KDETrees} \cite{KDE,KDETree}.}
    \label{fig:KDE_ROC}
\end{figure*}

To get the ROCs for the two samples, 
we conduct experiments described in Algorithm \ref{alg:sim1} with $r = 500$ and $g = 1000$.  More specifically, for each $R$, we take all $1000$ trees from $\mathbb{T}_1$ and we take one tree from $\mathbb{T}_2$.  Then we estimate probability distribution of gene trees using the tropical density estimator described in Equation \eqref{mastereq} (Fig. \ref{fig:KDE_ROC}, Left) and with {\tt KDETrees} (Fig. \ref{fig:KDE_ROC}, Right).  We iterate this process $500$ times.  Therefore, we have estimated probabilities for $1000$ trees in $\mathbb{T}_1$ and for $500$ trees in $\mathbb{T}_2$. 

In this next experiment (Fig. \ref{fig:KDE_ROC}, Right), we compare the results against {\tt KDETrees} from \cite{KDETree,KDE} with the Billera-Holmes-Vogtmann (BHV) metric \cite{BHV}.
We run computational experiments in MACPRO with 2.4 GHz 8-Core Intel Core i9 processor and 64 GB 2667 MHz DDR4 memory.  The computational time for one iteration with our tropical KDE is 9.54 seconds and with {\tt KDETrees} is 1.27 minutes.  

\begin{table}{Area Under the Curves (AUCs)}
    \centering
    \begin{tabular}{|c|cccccc|}\hline
    $R$ & 0.25 & 0.5 & 1 & 2 & 5 & 10\\\hline
         Tropical &$0.54$& $0.61$& $0.71$& $0.88$& $1.00$&  $1$\\
         BHV & $0.51$& $0.54$ & $0.54$& $0.72$& $0.98$ & $1$\\\hline
    \end{tabular}
    \caption{Area Under the Curves (AUCs) for the KDE with the tropical metric and the BHV metric via {\tt KDETrees}. }
    \label{tab:AUC}
\end{table}

\begin{table}{Apicomplexa gene sets identified as outliers by KDE with the tropical metric}
    \centering
    \begin{tabular}{|c|c|p{5cm}|}\hline
        \# & Gene ID & Function \\\hline
         691 & PFA0310c & calcium-transporting ATPase\\
         566 & PF13\_0257 & glutamate--tRNA ligase\\
         650 & PF11\_0358 & DNA-directed RNA polymerase, beta subunit, putative\\
         730& PFL0930w & clathrin heavy chain, putative\\
         615 & PF13\_0063 & 26S proteasome regulatory subunit 7, putative\\
         712 & MAL13P1.274 & serine/threonine protein phosphatase pfPp5\\
         630 & PFL2120w & hypothetical protein, conserved\\
         625 & PFD1090c & clathrin assembly protein, putative\\
         755& PF10\_0148 & hypothetical protein\\
         708 & PFC0140c & N-ethylmaleimide-sensitive fusion protein, putative\\
         497 & PF13\_0228 & 40S ribosomal subunit protein S6, putative\\
         690 & MAL8P1.134 & hypothetical protein, conserved\\
         503 & PF13\_0178 & translation initiation factor 6, putative\\\hline
    \end{tabular}
    \caption{Apicomplexa gene sets identified as outliers by KDE with the tropical metric.  All annotations except 728 are putative. Based on the gene set designations in \cite{kuo}. Gene set represented by GeneID for {\it P.falciparum}.} 
    \label{tab:outlyinggenes}
\end{table}

\subsection{Applications to Apicomplexa Data}

In this section we apply a tropical KDE with the HAR algorithm over the space of ultrametrics to the Apicomplexa dataset which consists of  268  orthologous sequences  with  eight  species  of  protozoa  from  \cite{kuo}.
There are eight species in each alignment in the set: {\it Babesia  bovis} (Bb), {\it Cryptosporidium
  parvum} (Cp), {\it Eimeria tenella} (Et) [15], {\it
  Plasmodium falciparum} (Pf) [11], {\it Plasmodium  vivax} (Pv),
{\it Theileria  annulata} (Ta),  and {\it Toxoplasma
  gondii} (Tg).  An outgroup is a free-living ciliate, {\it
  Tetrahymena  thermophila} (Tt).

The gene trees in the $0.05$ lower tail of the estimated distribution of gene trees using the tropical KDE are trees with their IDs 691, 566, 650, 730, 615, 712, 630, 625, 755, 708, 497, 690, 503 (ordered by the smallest probabilities to the largest).  Details of these outlying gene trees can be found in Table \ref{tab:outlyinggenes}. 

\begin{figure*}
    \centering
    \includegraphics[width=0.6\textwidth]{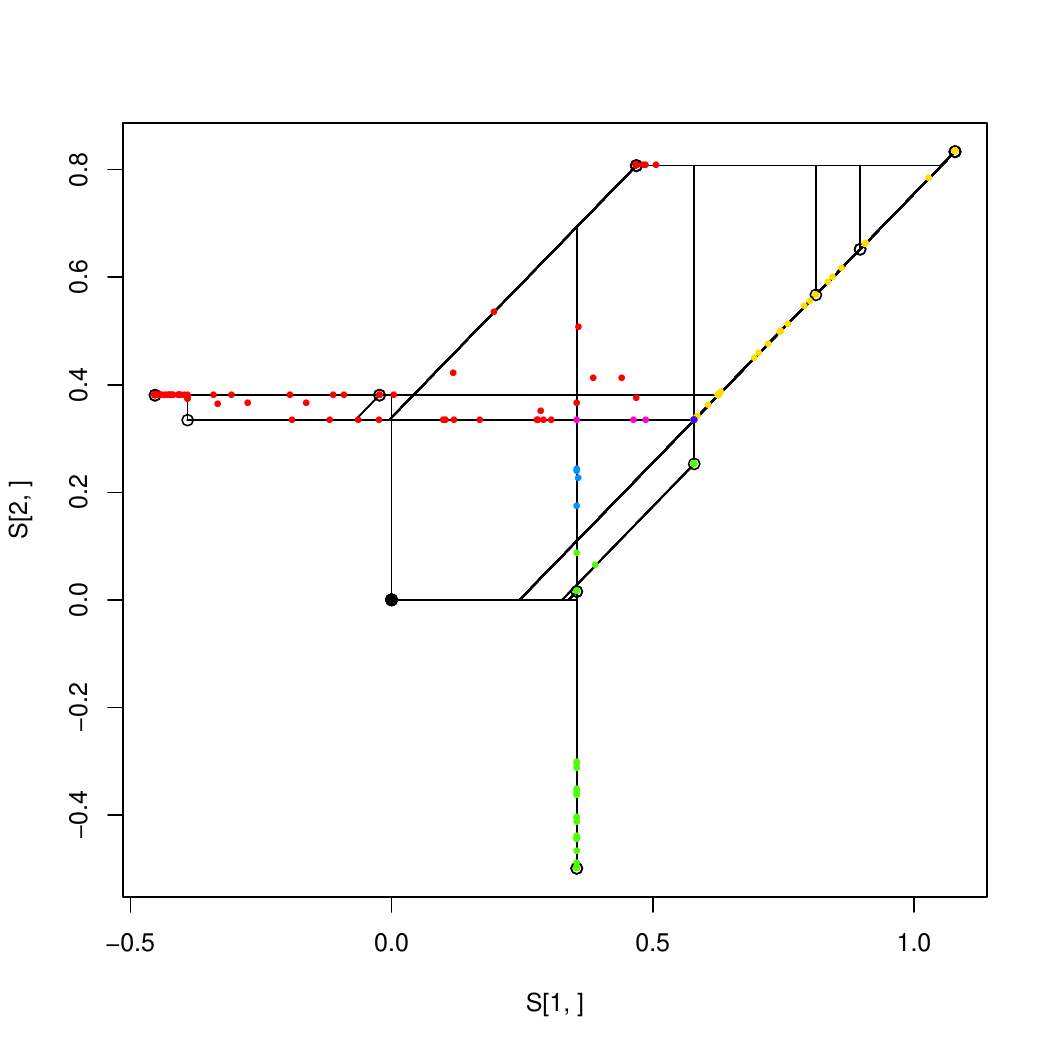} 
    \includegraphics[width=0.7\textwidth]{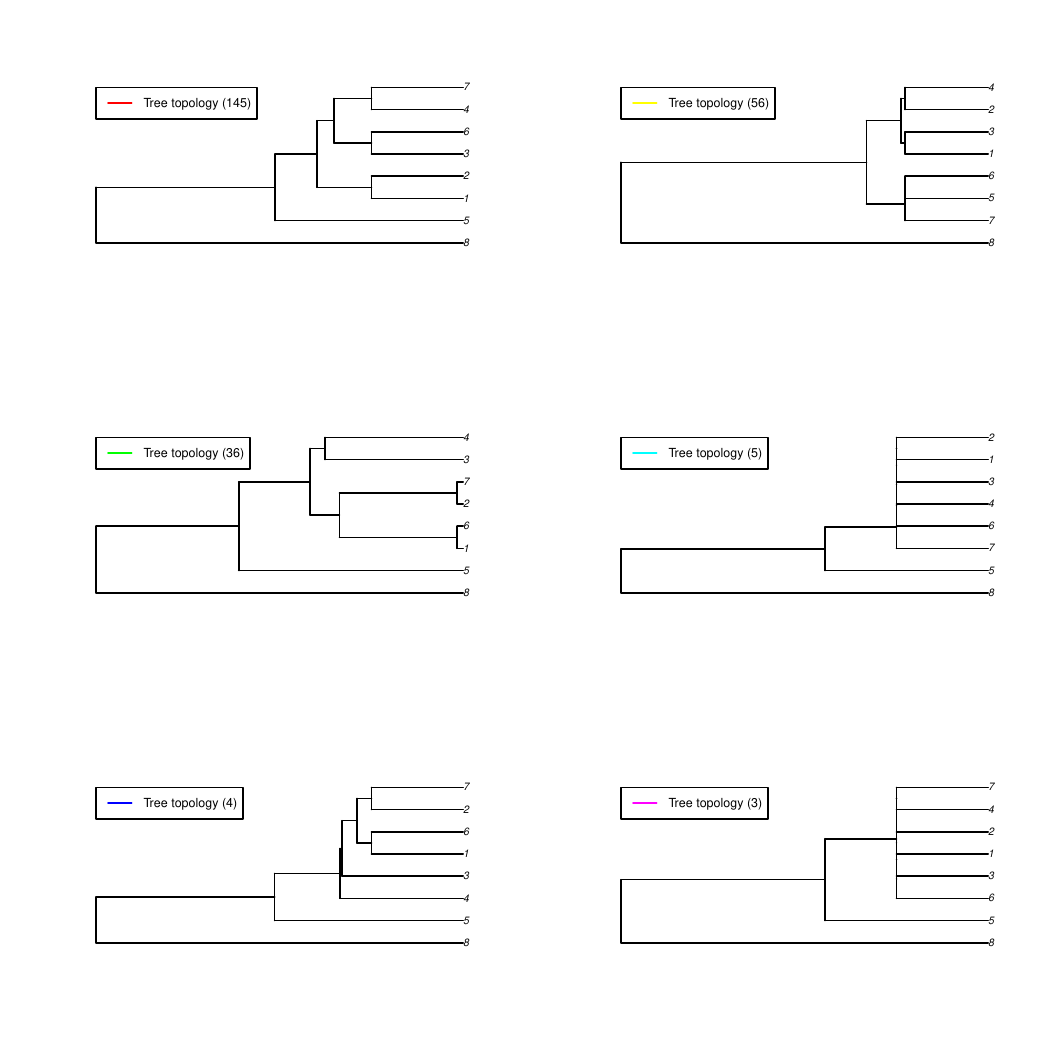} 
    \caption{Estimated tropical PCA developed by Yoshida et al.~in \cite{YZZ} using the Hit and Run algorithm developed by Yoshida et al.~\cite{YMB} with the Apicomplexa dataset from \cite{kuo}. In the figures above, a leaf label $1$ is for the label "Pv",  a leaf label $2$ is for the label "Pf", a leaf label $3$ is for the label "Tg", a leaf label $4$ is for the label "Et", a leaf label $5$ is for the label "Cp", a leaf label $6$ is for the label "Ta", a leaf label $7$ is for the label "Bb", a leaf label $8$ is for the label "Tt", the outgroup.  (Top) 1000 iterations were conducted for each vertex of the principal tropical triangle. The black points represent ``outlying'' gene trees.  (Bottom) the top $6$th tree topologies of the projected observations onto the best-fit tropical triangle after $1000$ iterations. The number inside of the parentheses for each tree topology is the number of projected trees which have the particular tree topology. The sum of residuals is $307.682$.}
    \label{fig:api}
\end{figure*}

In addition, we visualize the distribution of gene trees of Apicomplexa using tropical principal component analysis (PCA) developed by Yoshida et al.~\cite{YZZ} via the best-fitted tropical triangle, principal tropical polytope, shown in Fig.~\ref{fig:api}.  Suppose we have a $s \times e$ matrix $\mathcal{D}$ whose rows represent vertices of the best-fitted tropical polytope over $\mathbb{R}^e/\mathbb{R}{\bf 1}$ via the tropical PCA.  Recall that the tropical convex hull of the rows of $\mathcal{D}$ is isometric (linear translation) to the tropical convex hull of the columns of $\mathcal{D}$ \cite[Theorem 5.2.21]{MS}.  In our case we have $s = 3$ and $e = \binom{8}{2} = 28$.  Therefore, these unfilled circles in Fig.~\ref{fig:api} represent the columns of $\mathcal{D}$ whose rows are the vertices of the best-fitted tropical triangle for Apicomplexa data set from \cite{kuo}.
Filled black points in Fig.~\ref{fig:api} represent ``outlying'' gene trees.  The bottom of Fig.~\ref{fig:api} shows tree topologies which appear most frequently in the best-fit tropical triangle estimated.  The number next to each tree topology in the plot is the number of observations appear in the best-fit tropical triangle.




\section{Discussion}\label{sec:discussion}

\subsection{Simulation Study}

In \cite{KDE,KDETree}, Weyenberg et al.~showed that {\tt KDETrees} outperformed software {\tt Phylo-MCOA}.  In general it works well.  However, the biggest problem when using {\tt KDETrees} is that the normalizing constant, $C(T_i)$, of the function $k_{\rm BHV}(T, T_i)$ for all $T_i \in \mathcal{T}_m$ varies. However, using the tropical metric it seems that we do not have the same issue.
From the computation using the HAR sampler from \cite{YMB}, we estimate that $C$, the normalizing constant of $kK(K. T_i)$, is constant for all $T_i \in \Tn$ while with the BHV metric, the normalizing constant $C(T_i)$ for $kK_{rm BHV}(T, T_i)$ varies for $T_i \in \mathcal{T}_m$.  Therefore, we do not have to compute the normalizing constant for each observation in $\mathcal{S}$ when using the tropical metric in order to estimate the gene tree distribution from a sample $\mathcal{S} \subset \Tn$.  This is not the case when using the BHV metric, since the normalizing constant varies with each $T_i$, requiring computation for each sample. This makes the computational time for estimating the gene tree distribution much faster with the tropical metric as compared with using the BHV metric.  



Weyenberg et al. in~\cite{KDETree} estimate the normalizing constant $C(T_i)$ for each observation $T_i$ in a sample by using the {\em cone} distance between two trees on the BHV coordinates where one tree in the BHV coordinates goes through on the straight line to the origin (the star tree) and then goes through on the straight line to the other tree.  This can lead to large errors in the estimation and may affect the performance of {\tt KDETrees}.
As we can see from Table \ref{tab:AUC} and Figure \ref{fig:KDE_ROC}, our proposed non-parametric estimation of the gene tree distribution outperforms {\tt KDETrees} proposed by Weyenberg et al.~\cite{KDE,KDETree} for all $R = 0.25, 0.5, 1, 2, 5, 10$. 

\subsection{Apicomplexa}

In this section we summarize the analysis on outliers identified from the non-parametric estimation of gene tree distribution we propose in this research. Sequence alignments used to derive gene trees were judged to be poor if gene annotation errors were evident and likely reduced the accuracy of the alignment. Here we have
Pf = {\it Plasmodium falciparum}, Pv  = {\it Plasmodium vivax},  Bb = {\it Babesia  bovis},  Ta  = {\it Theileria annulata}, Et = {\it Eimeria tenella}, Tg = {\it Toxoplasma gondii},  Cp = {\it Cryptosporidium parvum}, and  Tt = {\it Tetrahymena thermophila} (outgroup).
\begin{itemize}
    \item {\bf PFA0310c}: Generally good alignment of sequences. The tree topology is mostly consistent with species phylogeny, except Tg and Et are clustered with the outgroup Tt rather than the expected Cp. 
    \item {\bf PF13\_0257}: Poor alignment in the N-terminal portion of the sequences. Long C-terminal extension in the outgroup Tt. There are several anomalies in the tree topology. The outgroup Tt clustered with the piroplasms Ta and Bb. The intestinal parasite Cp clustered with malaria parasites Pv and Pf. 
    \item {\bf PF11\_0358}: Good sequence alignment in blocks. Longer sequences for the malaria parasites Pf and Pv, including aN-terminal extension and several internal insertions. These potentially reflect incorrect gene annotation. Pf and Pv branch deeper than the Tt outgroup branch.
    \item {\bf PFL0930w}: Good sequence alignment in blocks, but with multiple assorted insertions in the gene for different taxa. The tree topology is inconsistent with phylogeny. The outgroup Tt branched internally and clustered with the coccidian parasites Tg and Et.  
    \item {\bf PF13\_0063}: Overall good sequence alignment. The protein horter sequence for Et is shorter. There is a 50 amino acid repetitive insertion in Et, possibly reflecting a gene annotation error. The tree is generally consistent with phylogeny. The intestinal parasite Cp is on the basal branch with outgroup Tt.
    \item {\bf MAL13P1.274}: Good alignment in the C-terminal half of the protein sequences. Inconsistent alignment in the N-terminal half with an approximately100 amino acid. extension in Pf and Pv. The outgroup Tt clustered with the malaria parasites Pf and Pv; otherwise, the tree topology is largely consistent with phylogeny.
    \item {\bf PFL2120w}: Poor sequence alignment, with multiple sequence insertions in different species. The intestinal parasite Cp clustered with the piroplasms Bb and Ta. The outgroup Tt clustered with the malaria parasites Pf and Pv.
    \item {\bf PFD1090c}: Good sequence alignment. There is a long N-terminal extension in Et with homopolymeric stretches, likely reflecting incorrect gene annotation. The tree is very inconsistent with phylogeny. The piroplasm Ta clustered with the intestinal parasite Cp. The piroplasm Bb clustered with the malaria parasites Pv and Pf. The outgroup Tt is located on an internal branch with the coccidian parasites Tg and Et.
    \item {\bf PF10\_0148}: Generally good alignment in the N-terminal half of the proteins. Insertion present in the gene sequences for Pf and Pv. There is an approximate 100 residue C-terminal extension in Tg. The tree topology is inconsistent with phylogeny. The intestinal parasite Cp clustered with the malaria parasites Pf and Pv. The outgroup Tt clustered with the coccidian parasites Tg and Et.
    \item {\bf PFC0140c}: Good alignment in the central portion of the gene sequences. The gene sequence for Et is much shorter. The coccidian parasite Tg clustered with the malaria parasites Pf and Pv.
    \item {\bf PF13\_0228}: The sequence for the outgroup Tt is much longer than all others with long N- terminal and C-terminal extensions. There is very good sequence alignment in blocks, but with lengthy insertions for the outgroup Tt, possibly reflecting incorrect annotation of the gene. The piroplasms Bb and Ta do not form a monophyletic taxon.
    \item {\bf MAL8P1.134}: Good alignment in blocks. There is a much longer sequence for the outgroup Tt. in the malaria parasites Pf and Pv share a sequence insertion. The tree has the malaria parasites Pf and Pv clustered with the coccidian parasites Tg and Et.
    \item {\bf PF13\_0178}: Good sequence alignment. The tree has Tg and Et branched as the basal taxa.
\end{itemize}
\section{Conclusion}\label{sec:conclusion}

From computational experiments, the tropical metric outperforms the BHV metric proposed by Weyenberg et al.~\cite{KDE,KDETree} when using this approach for a non-parametric estimation of the gene tree distribution in terms of accuracy and computational time. 
Therefore, we intend to extend our methods to large-scale codivergence studies that will describe the tree space encompassing such ancestral gene pools. After doing so, outliers in that the tree space will represent such events in genome evolution as gene duplications, lateral gene transfer between species, retention of ancestral polymorphisms by balancing selection, or accelerated evolution by neofunctionalization. Even phylogenetic codivergence of regions within enzyme sequences are of interest due to the possibility of module or domain shuffling in gene evolution. Outlier trees may represent erroneous gene models, correction of which can enhance genome annotations; or they may represent genes with unusual evolutionary histories caused by horizontal gene transfer, trans-species (ancient) polymorphisms, or accelerated evolution due to positive selection and neofunctionalization.

With the combination of visualization via tropical principal component analysis (PCA) developed by Yoshida et al.~\cite{YZZ}, we can see how gene trees in a given sample are distributed over the space of phylogenetic trees.   Fig.~\ref{fig:api} shows the visualization via the tropical PCA with annotations of outlying gene trees written in black.  From Fig.~\ref{fig:api}, it seems that all outlying gene trees are projected onto the same point in the two dimensional tropical triangle.  It is not clear whether this is a unique case or it happens often.  

It is well-known that if we reconstruct a phylogenetic tree from a concatenated alignment from gene alignments (for example, \cite{10.1080/10635150601146041}), an estimated phylogenetic tree is not statistically consistent.  This means that no matter how large an input alignment is, the reconstructed phylogenetic tree from the concatenated alignment might not be converging to the true tree.  However, without the methodology proposed in this research, we might be able to obtain an {\em interval estimation} of a phylogenetic tree from a set of gene trees instead of a point estimation of a tree based on a concatenated alignment from gene alignments.  

There are still some open problems.  For example, with the Billera-Holmes-Vogtmann metric \cite{KDETree}, the normalizing constant $C(T_i)$ for the function $k_{\rm BHV}(T, T_i)$ varies for $T_i \in \mathcal{T}_m$. 
While the geodesic between random two trees under the BHV metric over the tree space goes through the origin, i..e., the star tree, with positive probability \cite{OP}, Yoshida and Cox showed that under the tropical metric, the tropical line segment (geodesic under the tropical metric) between two random trees on the tree space does not go through the origin, the star tree, with probability one \cite{shelby} if $m \geq 5$. Therefore, for small trees with $m < 5$,   the normalizing constant for $k(T, T_i)$ with the tropical metric for $T_i \in \Tn$ might vary depending on their central location $T_i \in \Tn$. However, for $m \geq 5$,  the normalizing constant for $k(T, T_i)$ with the tropical metric for $T_i \in \Tn$ seems to be constant for any point in $\Tn$ as we see from the example \ref{eg:normalizedconst} for $m = 10$.  However, it is not proven mathematically.  Thus, we have the following conjecture:
\begin{conjecture}
The integration
\[
C(T_i) = \int_{\Tn} k(T, T_i) dT
\]
is constant for any fixed $T_i \in \Tn$ for $m \geq 5$.
\end{conjecture}

\section*{Funding}
RY and DB are partially funded by NSF DMS 1916037. KM is partially funded by JSPS KAKENHI 18K11485 and 22H02364.

\bibliographystyle{natbib}
\bibliography{refs}

\begin{thebibliography}{}

\bibitem[Akian {\em et~al.}(2011)Akian, Gaubert, Viorel, and Singer]{AGNS}
Akian, M.  {\em et~al.} (2011).
\newblock Best approximation in max-plus semimodules.
\newblock {\em Linear Algebra Appl.}, {\bf 435}, 3261--3296.

\bibitem[Ane {\em et~al.}(2007)Ane, Larget, Baum, Smith, and Rokas]{Ane2007}
Ane, C.  {\em et~al.} (2007).
\newblock Bayesian estimation of concordance among gene trees.
\newblock {\em Mol. Biol. Evol.}, {\bf 24}, 412--426.

\bibitem[Ardila and Klivans(2006)Ardila and Klivans]{AK}
Ardila, F. and Klivans, C.~J. (2006).
\newblock The bergman complex of a matroid and phylogenetic trees. journal of
  combinatorial theory.
\newblock {\em Series B\/}, {\bf 96}(1), 38--49.

\bibitem[Billera {\em et~al.}(2001)Billera, Holmes, and Vogtmann]{BHV}
Billera, L.  {\em et~al.} (2001).
\newblock Geometry of the space of phylogenetic trees.
\newblock {\em Adv Appl Math\/}, {\bf 27}(4), 733--767.

\bibitem[Buneman(1974)Buneman]{Buneman}
Buneman, P. (1974).
\newblock A note on the metric properties of trees.
\newblock {\em J. Combinatorial Theory Ser. B.}, {\bf 17}, 48--50.

\bibitem[Cohen {\em et~al.}(2004)Cohen, Gaubert, and Quadrat]{CGQ}
Cohen, G.  {\em et~al.} (2004).
\newblock Duality and separation theorems in idempotent semimodules.
\newblock {\em Linear Algebra Appl.}, {\bf 379}, 395--422.

\bibitem[Horner and Pesole(2004)Horner and Pesole]{Horner}
Horner, D. and Pesole, G. (2004).
\newblock Phylogenetic analyses: a brief introduction to methods and their
  application.
\newblock {\em Expert Rev. Mol. Diagn.}, pages 339--350.

\bibitem[Joswig(2021)Joswig]{ETC}
Joswig, M. (2021).
\newblock {\em Essentials of tropical combinatorics\/}.
\newblock Graduate Studies in Mathematics. American Mathematical Society,
  Providence, RI.

\bibitem[Kubatko and Degnan(2007)Kubatko and Degnan]{10.1080/10635150601146041}
Kubatko, L.~S. and Degnan, J.~H. (2007).
\newblock {Inconsistency of Phylogenetic Estimates from Concatenated Data under
  Coalescence}.
\newblock {\em Systematic Biology\/}, {\bf 56}(1), 17--24.

\bibitem[Kuo {\em et~al.}(2008)Kuo, Wares, and Kissinger]{kuo}
Kuo, C.  {\em et~al.} (2008).
\newblock The apicomplexan whole-genome phylogeny: An analysis of incongruence
  among gene trees.
\newblock {\em Mol Biol Evol\/}, {\bf 25}(12), 2689--2698.

\bibitem[Lin {\em et~al.}(2017)Lin, Sturmfels, Tang, and Yoshida]{LSTY}
Lin, B.  {\em et~al.} (2017).
\newblock Convexity in tree spaces.
\newblock {\em SIAM Discrete Math\/}, {\bf 3}, 2015--2038.

\bibitem[Liu and Pearl(2007)Liu and Pearl]{Liu2007}
Liu, L. and Pearl, D.~K. (2007).
\newblock Species trees from gene trees.
\newblock {\em Syst. Biol.}
\newblock in press.

\bibitem[Maclagan and Sturmfels(2015)Maclagan and Sturmfels]{MS}
Maclagan, D. and Sturmfels, B. (2015).
\newblock {\em Introduction to Tropical Geometry\/}, volume 161 of {\em
  Graduate Studies in Mathematics\/}.
\newblock Graduate Studies in Mathematics, 161, American Mathematical Society,
  Providence, RI.

\bibitem[Maddison and Maddison(2009)Maddison and Maddison]{mesquite}
Maddison, W.~P. and Maddison, D. (2009).
\newblock Mesquite: a modular system for evolutionary analysis. version 2.72.
\newblock Available at \url{http://mesquiteproject.org}.

\bibitem[Monod {\em et~al.}(2019)Monod, Lin, Kang, and Yoshida]{MLKY}
Monod, A.  {\em et~al.} (2019).
\newblock Tropical foundations for probability \& statistics on phylogenetic
  tree space.

\bibitem[Owen and Provan(2011)Owen and Provan]{OP}
Owen, M. and Provan, S. (2011).
\newblock A fast algorithm for computing geodesic distances in tree space.
\newblock {\em IEEE/ACM Trans.~Computational Biology and Bioinformatics\/},
  {\bf 8}, 2--13.

\bibitem[Page {\em et~al.}(2020)Page, Yoshida, and
  Zhang]{10.1093/bioinformatics/btaa564}
Page, R.  {\em et~al.} (2020).
\newblock {Tropical principal component analysis on the space of phylogenetic
  trees}.
\newblock {\em Bioinformatics\/}, {\bf 36}(17), 4590--4598.

\bibitem[Paradis {\em et~al.}(2004)Paradis, Claude, and Strimmer]{APE}
Paradis, E.  {\em et~al.} (2004).
\newblock A{PE}: analyses of phylogenetics and evolution in {R} language.
\newblock {\em Bioinformatics\/}, {\bf 20}, 289--290.

\bibitem[Rannala {\em et~al.}(2020)Rannala, Edwards, Leach{\'e}, and
  Yang]{rannala:hal-02535622}
Rannala, B.  {\em et~al.} (2020).
\newblock {The Multi-species Coalescent Model and Species Tree Inference}.
\newblock In C.~Scornavacca, F.~Delsuc, and N.~Galtier, editors, {\em
  {Phylogenetics in the Genomic Era}\/}, pages 3.3:1--3.3:21. {No commercial
  publisher | Authors open access book}.

\bibitem[Speyer and Sturmfels(2009)Speyer and Sturmfels]{SS}
Speyer, D. and Sturmfels, B. (2009).
\newblock Tropical mathematics.
\newblock {\em Mathematics Magazine\/}, {\bf 82}, 163--173.

\bibitem[Takahata and Nei(1990)Takahata and Nei]{Takahata}
Takahata, N. and Nei, M. (1990).
\newblock Allelic genealogy under overdominant and frequency-dependent
  selectionand polymorphism of major histocompatibility complex loci.
\newblock {\em Genetics\/}, {\bf 124}, 967--978.

\bibitem[Tukey(1977)Tukey]{Tukey}
Tukey, J. (1977).
\newblock {\em Exploratory Data Analysis\/}.
\newblock Addison-Wesley, Boston, MA.

\bibitem[Weyenberg {\em et~al.}(2014)Weyenberg, Huggins, Schardl, Howe, and
  Yoshida]{KDE}
Weyenberg, G.  {\em et~al.} (2014).
\newblock {kdetrees: non-parametric estimation of phylogenetic tree
  distributions}.
\newblock {\em Bioinformatics\/}, {\bf 30}(16), 2280--2287.

\bibitem[Weyenberg {\em et~al.}(2016)Weyenberg, Yoshida, and Howe]{KDETree}
Weyenberg, G.  {\em et~al.} (2016).
\newblock Normalizing kernels in the {B}illera-{H}olmes-{V}ogtmann treespace.
\newblock {\em IEEE ACM T. Comput. Bi.}, page doi:10.1109/TCBB.2016.2565475.

\bibitem[Yoshida and Cox(2022)Yoshida and Cox]{shelby}
Yoshida, R. and Cox, S. (2022).
\newblock Tree topologies along a tropical line segment.
\newblock {\em Vietnam Journal of Mathematics\/}, {\bf 50}, 395--419.

\bibitem[Yoshida {\em et~al.}(2019)Yoshida, Zhang, and Zhang]{YZZ}
Yoshida, R.  {\em et~al.} (2019).
\newblock Tropical principal component analysis and its application to
  phylogenetics.
\newblock {\em Bulletin of Mathematical Biology\/}, {\bf 81}, 568--597.

\bibitem[Yoshida {\em et~al.}(2023)Yoshida, Miura, and Barnhill]{YMB}
Yoshida, R.  {\em et~al.} (2023).
\newblock Hit and run sampler from tropically convex sets.
\newblock {\em Algebraic Statistics\/}.
\newblock To appear.

\end{thebibliography}









\end{document}